\documentstyle[12pt,epsfig,subfigure,amssymb]{article}
\newcommand{\pp}{{\sf p}}                  
\newcommand{\uq}{{\sf u}}                  
\newcommand{\dq}{{\sf d}}                  
\newcommand{\sq}{{\sf s}}                  
\newcommand{\qq}{{\sf q}}                  
\def\la{{$\Lambda$}}
\def\al{{$\overline{\Lambda}$}}
\def\lal{{$\Lambda / \overline{\Lambda}$}}

\begin{document}
\begin{center}
\vspace{2cm}
{ \bf LAMBDA POLARIZATION IN LEPTON INDUCED REACTIONS} \\
\vspace{1.5cm}

{\large A.M.~Kotzinian}\\
\vspace{1cm}
{\it JINR, Dubna and YerPhI, Yerevan}
\vspace{3cm}
\begin{abstract}

Different phenomenological approaches for $\Lambda$ and $\overline{\Lambda}$
polarization in polarized semi-inclusive deep inelastic scattering and
electron-positron annihilation at $Z^0$ pole are considered. 
Current and future experiments will soon provide accurate enough data 
to study spin phenomena in these reactions 
and distinguish between various models.
\end{abstract}

\vspace{3cm}

Talk given at the \\
\vspace*{8pt}
 SPIN-97 VII Workshop on\\
\vspace*{3pt}
{\it  High Energy Spin Physics }\\
\vspace*{3pt}
7-12 July 1997, Dubna, Russia. \\
\vspace*{8pt}
{\it (to be published in the proceedings)}
\end{center}

\newpage

$\;\;\quad$  
The self-analysing properties of the $\Lambda (\overline{\Lambda})$ 
make this particle particularly suited for spin physics.
Non-trivial {\it negative} longitudinal polarization of $\Lambda$'s
produced in the target fragmentation region of the deep-inelastic 
$\bar \nu$ scattering, measured with
respect to the direction of the momentum transfer from the beam,
has been observed in WA59 experiments~\cite{wa59}. These data can be 
interpreted in a simple way~\cite{ekk96} by using the model of 
polarized intrinsic strangeness in polarized nucleon~\cite{ekks95}. 
According to this model a valence quark core
with (essentially) the na\"{\i}ve quark model spin content may be
accompanied by a spin--triplet \sq${\sf \bar s}$ pair in which the 
${\sf \bar s}$
antiquark is supposed to be negatively polarized, motivated by chiral
dynamics, and likewise the \sq~quark, motivated by $^3P_0$
quark condensation in the vacuum.

The essence of our argument~\cite{ekk96} is that the right-handed polarization
of the $\bar \nu$ beam is transferred to the hadrons via polarized
$W^-$-exchange, which selects preferentially one {\it longitudinal}
polarization state of the nucleon target.
Specifically, in most interactions
the $\bar \nu$-induced $W$ removes a {\it positively}-polarized
\uq~quark from the nucleon target, as seen in Fig. 1.

\begin{figure}[h]
\vspace*{-8mm}
\begin{center}
\mbox{ \psfig{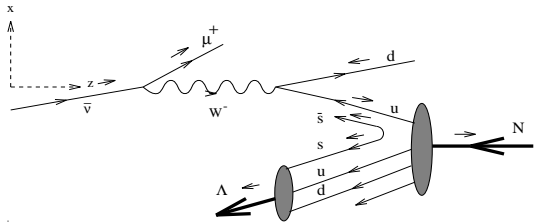} }
\end{center}
\vspace*{-9mm}
\caption{\it Dominant diagram for $\Lambda$ production in the
target fragmentation region due to scattering on a valence \uq~quark.
Each small arrow represent the longitudinal polarization
of the corresponding particle.}
\label{fig:uquark}
\vspace*{-2mm}
\end{figure}

In the na\"{\i}ve quark-parton model of deep-inelastic
$\nu$ or $\bar \nu$ scattering, the net longitudinal
polarization of a remnant \sq~quark , $P_{s}$, is given by
\begin{equation}
P_{s}=\frac{\sum_q c_{s\,q}N_q -
\sum_{\bar q} c_{s\,\bar q}N_{\bar q}}{N_{tot}},
\label{ps}
\end{equation}
where  $c_{s\,q}$ is the remnant \sq~quark spin-correlation
coefficient with the struck quark \qq,
$N_q$ ($N_{\bar q}$) is the total number of events selected
in which a quark (antiquark) is struck, and $N_{tot}=N_q+N_{\bar q}$
is the total number of events selected. 
According to the polarized strangeness model,
the polarization of the remnant \sq~quark is 100\%
{\it anticorrelated} with that of the valence quark, and 100\%
{\it correlated} with that of struck sea $\sf \bar s$ antiquark (see Fig. 2):

\begin{figure}[h]
\vspace*{-9mm}
\begin{center}
\mbox{ \psfig{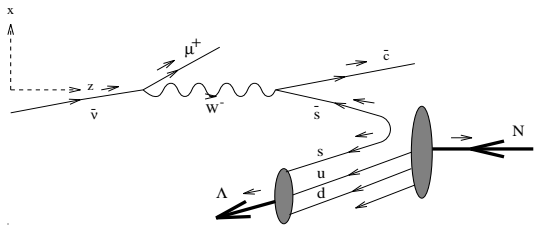} }
\end{center}
\vspace*{-8mm}
\caption{\it Diagram for $\Lambda$ production in a $\bar \nu \, N$
event due to W interaction with a $ \sf \bar s$ quark from the sea. As in
Fig. 1, the small arrows represent longitudinal polarizations. }
\label{fig:cquark}
\vspace*{-10mm}
\end{figure}

\begin{eqnarray}
\label{ccss}
c_{s\,u_{val}} &=& c_{s\,d_{val}}=-1,\nonumber\\
c_{s\,\bar q_{sea}} &=& \delta_{\bar s,\bar q}.\
\end{eqnarray}

In the simple quark model
the polarization of a directly-produced $\Lambda$ is the same as that of
the remnant \sq~quark. However, final-state $\Lambda$'s may also be
produced indirectly via the decays of heavier hyperon resonances,
which tends to dilute the $\Lambda$ polarization by a factor we denote
by $D_F$. Thus the final-state longitudinal $\Lambda$ polarization is
\begin{equation}
P_{\Lambda}=D_F P_{s},
\label{plam}
\end{equation}
The fraction of $\Lambda$'s
produced indirectly may vary with the kinematical conditions, e.g., it
may be higher when the invariant mass of the produced hadron
system is larger.
\vskip0.3cm

We have used the Lund Monte Carlo program LEPTO
to obtain numerical results.
In the Table 1 our results for the
polarization $P_s$ of the remnant \sq~quark in various ranges of
Bjorken $x$, together with the corresponding values of $P_{\Lambda}$
measured in WA59 experiment. We also tabulate the corresponding values
of $D_F$ inferred from our calculated values of $P_s$ and the measured
values of $P_{\Lambda}$.

\begin{table}[htb]
\begin{center}
\begin{tabular}{|c||c|c|c|} \hline
$x$ range      & $0 < x < 1$      & $0 < x < 0.2$    & $0.2 < x < 1$    \\
\hline \hline
$P_{\Lambda}$ in WA59 experiment
& $-0.63 \pm 0.13$ & $-0.46 \pm 0.19$ & $-0.85 \pm 0.19$ \\
\hline
$P_{s}$ in our model
& $-0.86$           & $-0.84$           & $-0.94$           \\
\hline
Dilution factor $D_F$
& $ 0.73 \pm 0.15$ & $ 0.55 \pm 0.23$ & $ 0.90 \pm 0.20$\\
\hline
\end{tabular}
\end{center}
\caption{\it $\Lambda$ polarization in the target fragmentation
region ($x_F < 0$).}
\label{tab:antineutrino}
\end{table}
\vskip0.3cm

The similar value and sign of $\Lambda$~polarization is expected also
for the $\nu$-beam.

It is interesting to contrast the above predictions with the expectation 
for the meson cloud model of DIS ~\cite{melth95}.
In such a model the $\Lambda$~ polarisation in the target fragmentation
region is expected to be zero for unpolarized target 
(in contradiction with WA59 data) and very strongly anticorrelated
with the target polarization. In the similar light-cone meson-baryon
fluctuation model \cite{bm96} one expect the negative polarization for
the produced $\Lambda$~ and zero (or slightly positive) 
for the produced $\overline{\Lambda}$~
whereas in our approach both polarization are expected to be negative.

Next we apply our model to predict
the polarization of $\Lambda$'s produced in the target fragmentation
region in the deep-inelastic scattering of polarized muons (electrons)
on both unpolarized and polarized nucleon targets.
It is easy to find the following expression for
the polarization of the remnant \sq~quark
\begin{equation}
P_{s_{rem}}
=\frac {\sum_q\,e_q^2[P_T\Delta q(x)-P_BD(y)q(x)]c_{s\,q}}
{\sum_q\,e_q^2[q(x)-P_BP_TD(y)\Delta q(x)]},
\label{psr}
\end{equation}
where $P_B$ and $P_T$ are the beam and target longitudinal
polarizations, $e_q$ is the quark charge,
$q(x)$ and $\Delta q(x)$ are the unpolarized 
and polarized quark distribution functions, and
$D(y)=[1-(1-y)^2]/[1+(1-y)^2]$.

The results of our calculations for a $\mu$ beam with the
longitudinal polarization $P_{\mu}=-0.8$ are shown in Fig. 6,
together with the cases $P_{\mu}=0$ and 0.8.

\begin{figure}[h]
\vspace*{-10mm}
\begin{center}
\mbox{\begin{tabular}[t]{cc}
\subfigure[{\it Proton target}]
{\psfig{file=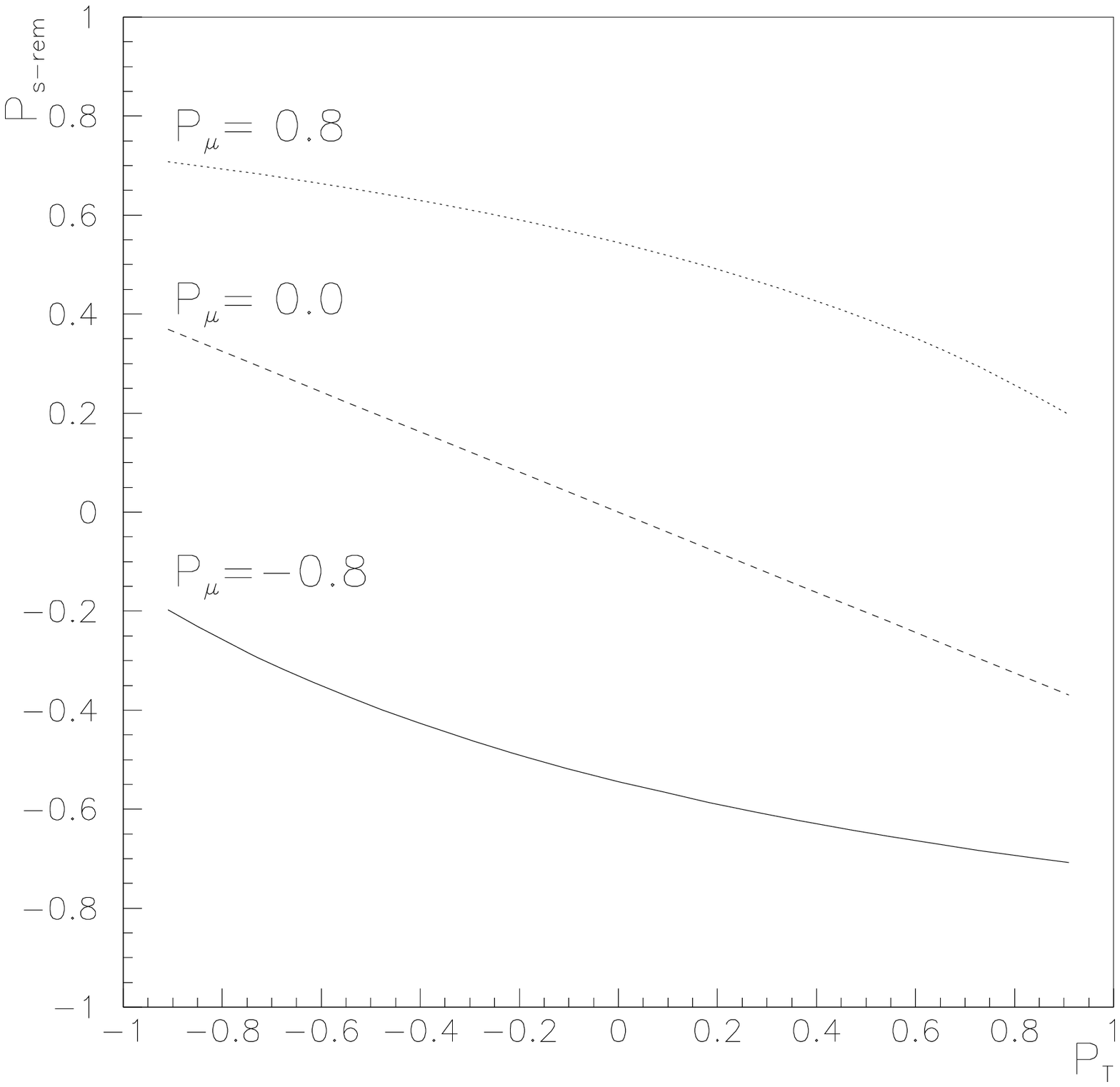,height=4cm,width=.35\textwidth}} &
\subfigure[{\it Ammonia target}]
{\psfig{file=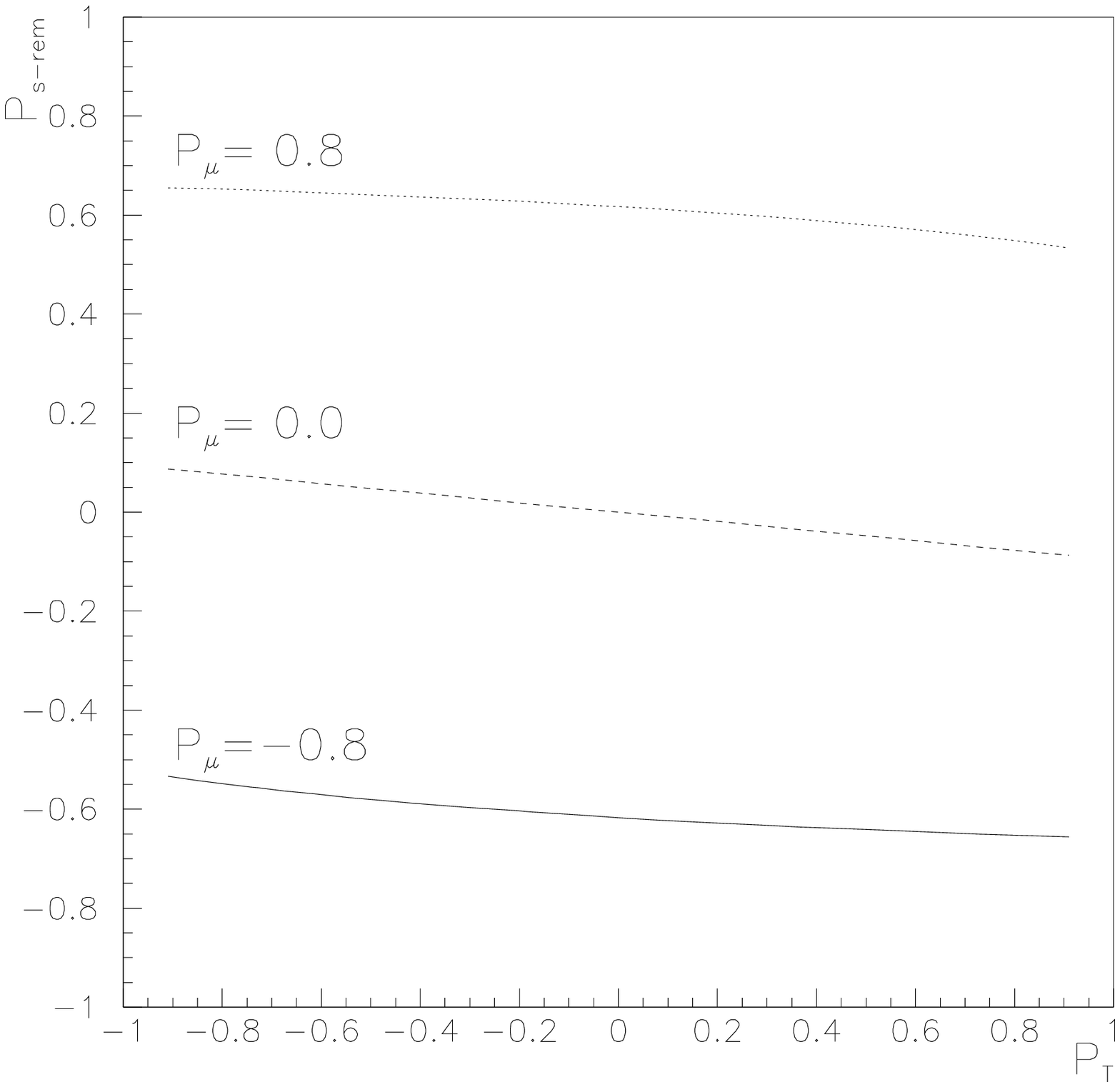,height=4cm,width=.35\textwidth}}
\end{tabular}}
\end{center}
\vspace*{-10mm}
\caption{\it Polarization of remnant \sq~quark for deep-inelastic
$\mu$ scattering, as a function of the target polarization $P_T$ for
different values of the beam polarization $P_{\mu}$.
a) for a proton target, b)for an ammonia target. We assume $E_{\mu}$=190
GeV  and the following cuts
were applied: $-0.3<x_F<0$, $x>0.15$, $0.5<y<0.9$.}
\label{fig:psmu}
\end{figure}
\vskip0.3cm

The produced $\Lambda$ polarization is given by equation (\ref{plam}).
We concluded from our analysis of the WA59 data that
in the valence-quark region
the fragmentation dilution factor $D_F \gtrsim 0.7$. Therefore, we
expect large polarization effects also for $\Lambda$
production in the target fragmentation region in deep-inelastic
$\mu \,N$ scattering.
\vskip0.3cm

Complementary nonperturbative phenomenon to polarized parton
distribution in the polarized nucleon is the polarization 
transfer in the quark fragmentation process. 
I will present the results of our calculations~\cite{kbh97,comp} for 
different phenomenological spin transfer mechanisms 
for the $\Lambda$~and $\bar{\Lambda}$~
longitudinal polarization 
in various lepton induced processes.

The leading twist unpolarized ($D_q^\Lambda (z)$)
and polarized ($\Delta D_q^\Lambda (z)$) quark fragmentation
functions to a \la~hyperon are defined as:

\begin{eqnarray}
D_q^{\Lambda}(z)&=& D_q^{+ \;\Lambda}(z)+ D_q^{- \;\Lambda}(z) \\
\Delta D_q^{\Lambda}(z)&=& D_q^{+\;\Lambda}(z)- D_q^{-\;\Lambda}(z),
\end{eqnarray}
where  $D_q^{+ \;\Lambda} (z)$ ($D_q^{- \;\Lambda} (z)$) is the spin
dependent quark fragmentation functions for the \la~spin parallel 
(antiparallel) to that of the initial quark $q$, and $z$
is the quark energy fraction carried by the $\Lambda$ hyperon.

We parametrize the polarized quark fragmentation functions as 

\begin{equation}
\Delta D_q^\Lambda (z) = C_q^\Lambda (z) \cdot D_q^\Lambda (z),
\end{equation}
where $C_q^\Lambda (z)$ are the spin transfer coefficients.
 
Two different descriptions of the spin transfer mechanism in 
the quark fragmentation to a \lal~hyperon are considered.
The first one is based on the non-relativistic quark model SU(6)
wave functions, where the \la~spin is carried only by its constituent
\sq~quark.
Therefore, the polarization of directly produced \la's is determined
by that of the \sq~quark only, while \la's coming from decays of heavier
hyperons inherit a fraction of the parent's polarization, which
might originate also from other quark flavors
(namely \uq~and \dq).
In this scheme the spin transfer is discussed in terms of
{\it constituent quarks}.
Table~\ref{tab:bgh} shows the spin transfer coefficients
$C_q^\Lambda$ for this case~\cite{bi,gh}.
A particular case is given by a simpler assumption that the \la~hyperon
gets its polarization from \sq~quarks only.
In the following we will refer to the former description as $BGH$
({\it for} Bigi, Gustafson, and H\"akkinen)
and the latter as $NQM$ ({\it for} na\"{\i}ve quark model).

\begin{table}
\begin{center}
\begin{tabular}{|c||c|c|c|c|}
\hline
\la's parent & $C_u^\Lambda$ & $C_d^\Lambda$ & $C_s^\Lambda$ & 
$C_{\bar q}^\Lambda$\\ \hline \hline
Quark          & $0$    & $0$    & $+1$   & $0$ \\ \hline 
$\Sigma^0$     & $-2/9$ & $-2/9$ & $+1/9$ & $0$ \\ \hline 
$\Sigma(1385)$ & $+5/9$ & $+5/9$ & $+5/9$ & $0$ \\ \hline
$\Xi$          & $-0.3$ & $-0.3$ & $+0.6$ & $0$ \\ \hline
\end{tabular}
\end{center}
\vspace*{-3mm}
\caption{\it Spin transfer coefficients according to non-relativistic SU(6)
quark model.}
\label{tab:bgh}
\vspace*{-2mm}
\end{table}

The second approach is based on the $g_1^\Lambda$ {\it sum rule} for the
first moment of the polarized quark distribution functions in
a polarized \la~hyperon, which was derived by Burkardt and Jaffe~\cite{BuJ93}
in the same fashion as for the proton one ($g_1^{\pp}$).
We assume that the spin transfer from a polarized quark \qq~to
a \la~is proportional to the \la~spin carried by that flavor,
{\it i.e.} to $g_1^\Lambda$.
Table~\ref{tab:bjsr} contains the spin transfer coefficients $C_q^\Lambda$,
which were evaluated using the experimental values for $g_1^{\pp}$.
Two cases are considered~\cite{Jaf96}:
in the first one only valence quarks are polarized;
in the second case also sea quarks and antiquarks
contribute to the \la~spin.
In the following we will refer to the first one as {\it BJ-I} and the
second one as {\it BJ-II}.

\begin{table}
\begin{center}
\begin{tabular}{|c||c|c|c|c|}
\hline
  & $C_u^\Lambda$ & $C_d^\Lambda$ & $C_s^\Lambda$ & $C_{\bar q}^\Lambda$ \\ 
\hline \hline
{\it BJ-I}    & $-0.20$  & $-0.20$  & $+0.60$ & $0.0$  \\ \hline
{\it BJ-II}   & $-0.14$  & $-0.14$  & $+0.66$ & $-0.06$  \\ \hline
\end{tabular}
\end{center}
\vspace*{-3mm}
\caption{\it Spin transfer coefficients according to the Burkardt-Jaffe 
$g_1^\Lambda$ sum rule.}
\label{tab:bjsr}
\vspace*{-2mm}
\end{table}

In the $g_1^\Lambda$ {\it sum rule} scheme a negative spin transfer from
\uq~and \dq~quarks to a \la~hyperon is predicted.
This effect can be understood qualitatively even if the spin
of the \la~is determined by its constituent \sq~quark only:
in some cases the fragmenting \uq~or \dq~quark will
become a sea quark of the constituent \sq~quark, and the spin
of the constituent \sq~quark will be anti-correlated to the spin of
the fragmenting quark~\cite{ekks95,ekk96}.
Another possibility occurs when the \la~is produced as a second rank
particle in the fragmentation of a \uq~or \dq~quark.
If the first rank particle was a pseudoscalar strange meson,
then the spin of the ${\sf \bar s}$ antiquark has to be opposite to that
of the \uq~(\dq) quark,
and since the \sq${\sf \bar s}$ pair created from the vacuum in the string
breaking is assumed to be in a $^3P_0$ state,
the \sq~quark is also oppositely polarized to the \uq~or \dq~quark.
This last mechanism of the spin transfer can be checked by measuring the 
\la~polarization for a sample of events containing fast $K$ mesons.

For charged lepton DIS the fragmenting quark longitudinal polarization
is given by the simple parton model expression~\cite{comp,ekk96}

\begin{equation}
P_{q'} \, (x,y) =
\frac{P_B D(y) q(x) + P_T \Delta q(x)}
{q(x) + P_B D(y) P_T \Delta q(x)}.
\label{eq:lambdap}
\end{equation}

For neutrino scattering the flavor changing charged current weak interaction
selects left-handed quarks (right-handed antiquarks),
giving 100~\% polarized fragmenting quarks. 
The Standard Model predicts a high degree of longitudinal polarizations
for quarks and antiquarks produced in $Z^0$ decays: 
$P_s=P_d=-0.91, \;P_u=P_c=-0.67$.

\begin{figure}[h!]
\vspace*{-10mm}
\begin{center}
\mbox{\epsfxsize=10cm\epsffile{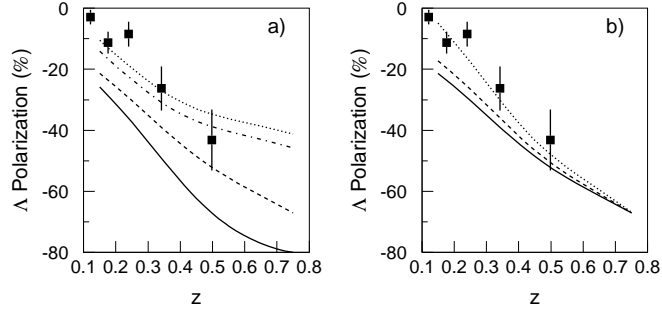}}
\end{center}
\vspace*{-11mm}
\caption{\it a) \lal~polarization at the $Z^0$ pole for different mechanisms
of spin transfer: solid line - $NQM$, dashed - $BGH$, dotted - {\it BJ-I},
and dot-dashed - {\it BJ-II}.
The experimental data (full squares) are from~\protect\cite{aleph}.
b) comparison between predictions using the {\it BGH} model
for the \la~polarization in our analysis
(solid line) and the analysis of~\protect\cite{aleph} assuming that
only \sq~quarks contribute to \la~polarization (dashed), and additionally
that only first rank \la's inherit a fraction of the fragmenting quark
polarization (dotted).}
\label{fig:lamz0pol}
\end{figure}

\begin{figure}[h!]
\vspace*{-15mm}
\begin{center}
\mbox{\epsfxsize=10cm\epsffile{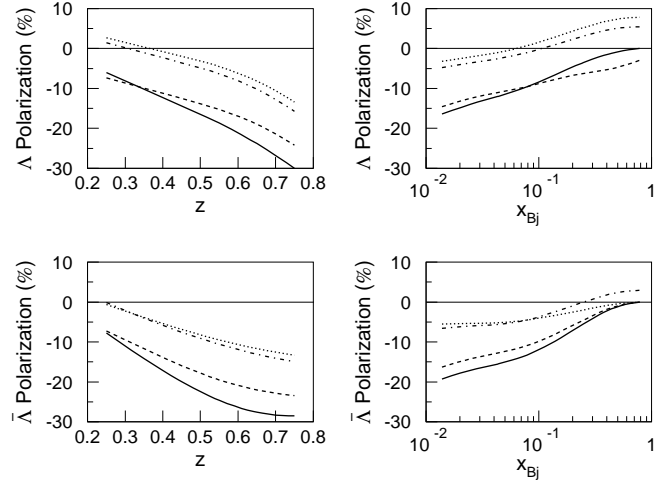}}
\end{center}
\vspace*{-11mm}
\caption{\it \la~and \al~longitudinal polarization in the current 
fragmentation region for DIS of polarized $\mu^+$'s on an unpolarized
target for different mechanisms of spin transfer:
solid line - $NQM$, dashed - $BGH$, dotted - {\it BJ-I}, and
dot-dashed - {\it BJ-II}.}
\label{fig:unpolarized}
\end{figure} 

\begin{figure}[h!]
\vspace*{-10mm}
\begin{center}
\mbox{\epsfxsize=10cm\epsffile{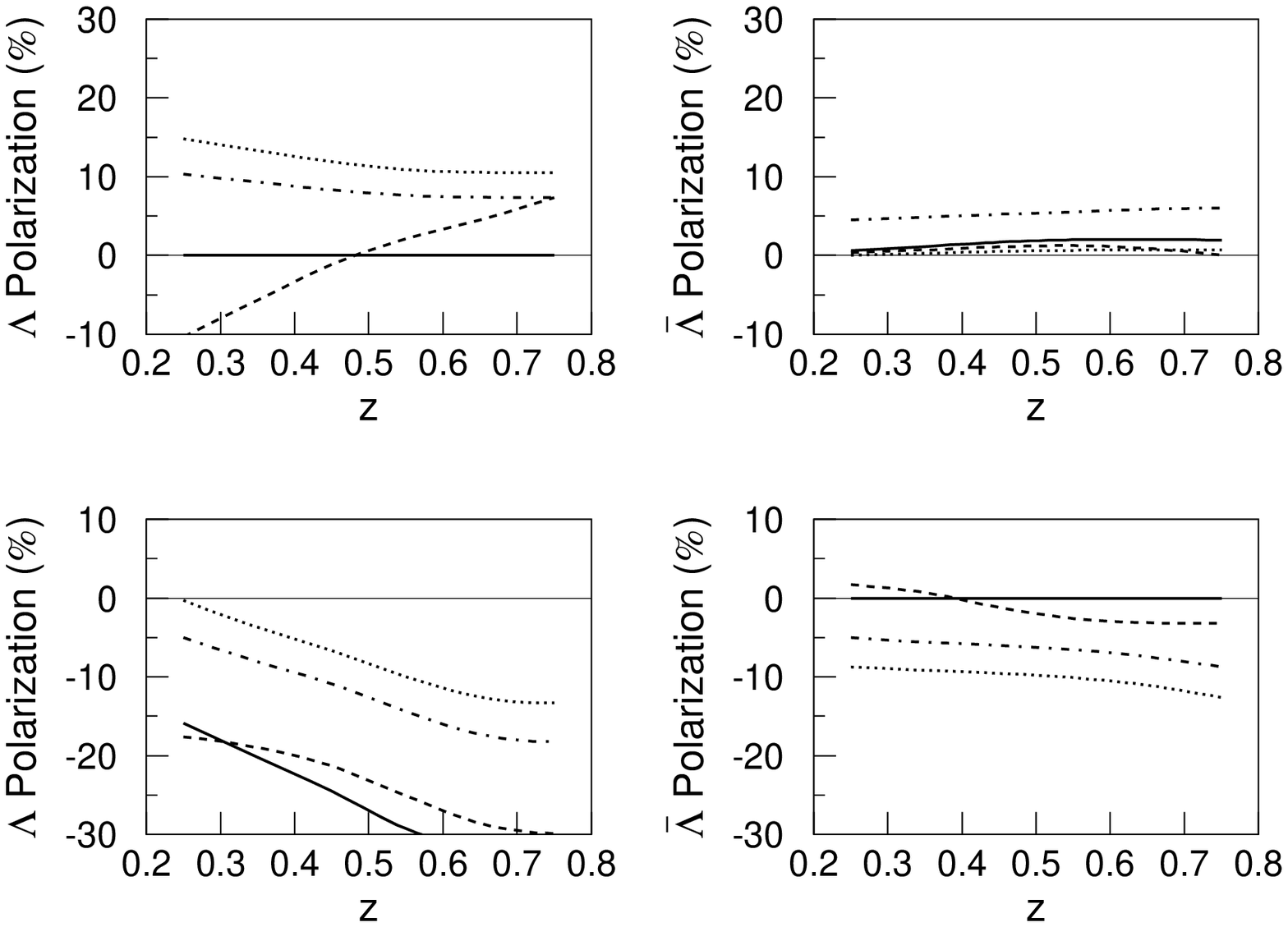}}
\end{center}
\vspace*{-11mm}
\caption{\it \lal~polarization in the current fragmentation region in 
$\nu$--DIS (upper plots) and $\overline{\nu}$--DIS (lower plots):
solid line - $NQM$, dashed - $BGH$, dotted - {\it BJ-I}, and
dot-dashed - {\it BJ-II}.}
\label{fig:neutrino}
\end{figure} 

The results~\cite{kbh97}
for longitudinal \lal~polarization in different processes are
presented in the Figs.~\ref{fig:lamz0pol}, \ref{fig:unpolarized} and 
\ref{fig:neutrino}.  As one can see from the  Fig.~\ref{fig:lamz0pol}
the existing data~\cite{aleph} does not allow to distinguish between 
$BGH$ and $BJ$ mechanisms for the spin transfer. But, as one can see
from Figs.~\ref{fig:unpolarized} and \ref{fig:neutrino} in the current
fragmentation region of deep inelastic scattering the predicted 
\lal~polarizations are rather different for the two mechanisms. 
Our studies have shown that the \lal~polarization in the current
fragmentation region of polarized electro-production 
is less sensitive to the target polarization (in general)
and especially to $\Delta {\sf s}$. 

The new accurate data which 
are soon expected from NOMAD~\cite{nomad}, HERMES~\cite{herm} and
COMPASS~\cite{comp} experiments will provide possibility for
detailed study of these interesting phenomena.

\end{document}